\documentclass[aps,amsmath,prb,twocolumn,a4paper,showpacs]{revtex4}
\usepackage{graphicx}
\begin{document}

\newcommand{\figureheight}{8.2 cm}
\newcommand{\putfig}[2]{\begin{figure}[h]
\special{isoscale #1.bmp, \the\hsize \figureheight}
\vspace{\figureheight} \caption{#2} \label{fig:#1}
\end{figure}}

\newcommand{\eqn}[1]{(\ref{#1})}
\newcommand{\be}{\begin{equation}}
\newcommand{\ee}{\end{equation}}
\newcommand{\bea}{\begin{eqnarray}}
\newcommand{\eea}{\end{eqnarray}}
\newcommand{\bean}{\begin{eqnarray*}}
\newcommand{\eean}{\end{eqnarray*}}
\newcommand{\nn}{\nonumber}

\title{ Transport through a double barrier for interacting
quasi one-dimensional electrons in a Quantum Wire in  the presence
of a transverse magnetic field }
\author{S. Bellucci $^1$ and P. Onorato $^1$ $^2$ \\}
\address{
$^1$INFN, Laboratori Nazionali di Frascati,
P.O. Box 13, 00044 Frascati, Italy. \\
$^2$Dipartimento di Scienze Fisiche, Universit\`{a} di Roma Tre,
Via della Vasca Navale 84, 00146 Roma, Italy}
\date{\today}
\begin{abstract}
We discuss the Luttinger Liquid behaviour of a semiconducting
Quantum Wire. We show that the measured value of the  bulk
critical exponent, $\alpha_{bulk}$,  for the tunneling density of
states can be easily calculated.

Then, the problem of the transport  through a Quantum Dot formed
by two Quantum Point Contacts along the Quantum Wire, weakly
coupled to spinless Tomonaga-Luttinger liquids is studied,
including the action of a strong transverse magnetic field $B$.
 The known magnetic dependent peaks of the
conductance, $G(B)$, in the ballistic regime at a very low
temperature, $T$, have to be reflected also in the transport at
higher $T$ and in different regimes. The temperature dependence of
the maximum $G_{max}$ of the conductance peak, according to the
Correlated Sequential Tunneling theory, yields the power law
$G_{max}\propto  T^{2\alpha_{end}-1}$, with the critical exponent,
$\alpha_{end}$, strongly reduced by $B$.

This behaviour suggests the use of a similar device as a magnetic
field modulated transistor.

\end{abstract}

\maketitle

\section{Introduction}

Progress in semiconductor device fabrication and carbon
technology allowed for the construction of several low-dimensional
structures at the nanometric scale, and many novel transport
phenomena have been revealed.


The electron-electron (e-e) correlation effects, usually  negligible in
three-dimensional devices,  attract considerable interest, because
of the dominant role which they play in one dimension, by
determining   the physical properties of a one dimensional (1D)
metal.

The main consequence of the e-e Coulomb repulsive
interaction  in 1D systems of interacting electrons is the
formation of a Tomonaga-Luttinger  liquid (TLl) with properties
 that  are dramatically different from the ones  of usual
metals with a Fermi liquid of electrons\cite{egg,TL,TLreviwew}.

{Because of the e-e interaction, in  the TLl
Landau quasiparticles are unstable and the low-energy excitation is
achieved by exciting an infinite number of plasmons (collective
electron-hole pair modes), making the transport intrinsically
different from that of a Fermi liquid. Hence, it follows a power-law
dependence of physical quantities, such as the tunneling density
of states (TDOS), as a function of the energy or the temperature.}

\

{\it Transport in 1D -}
Thus, the transport through 1D devices attracts considerable
interest,
 because it displays a power-law zero-bias
anomaly (ZBA) for the conduction. The tunneling conductance, $G$,
reflects the power law dependence of the DOS in a small bias
experiment\cite{kf} \bea \label{gbul}
 G=dI/dV\propto
T^{\alpha_{bulk}}, \eea for $eV_b\ll k_BT$, where $V_b$ is the bias
voltage, $T$ is the temperature and $k_B$ is Boltzmann's constant.
Many theoretical works and experiments, during  the last
 decade,
 concentrated on the power-law behavior of the
electron tunneling by analyzing quantum Hall edge systems
\cite{kf,wen,fn}, carbon nanotubes (CNs)\cite{cnts,tubes}, and
semiconductor Quantum Wires (QWs)\cite{tarucha,yacoby}.

The bulk critical exponent can be obtained in several different
ways\cite{TLreviwew} and has the form
\begin{equation}\label{al1}
  \alpha_{bulk}=\frac{1}{4} \left(K+\frac{1}{K}-2 \right).
\end{equation}
If we follow the RG approach \cite{noi,noia1,noia2}  for the
unscreened e-e interaction we obtain \bea\label{k} \sqrt{1 +
\frac{U_0(q_c, B)}{ (2 \pi  {v}_F)}} =\frac{1}{K}, \eea where
$v_F$ is the Fermi velocity and $ U_0(p,B) $ corresponds to the
Fourier transform of the 1D  e-e interaction potential, also
depending on the magnetic field $B$.
 Thus
$K$ is a function of the interaction strength ($K<1$ corresponding
to repulsive interaction) while $q_c=2\pi/L$ is the natural
infrared cut-off depending on the longitudinal length of the quasi
1D device.

\

The power-law behaviour characterizes also the thermal dependence
of $G$ when an impurity is present along the 1D devices. The
theoretical approach to the presence of obstacles  mixes two
theories corresponding to  the single particle scattering and
the TLl theory of interacting electrons. In fact the presence of a
barrier is usually  modeled by a potential barrier $V_B({\bf r})$
and  the single particle scattering gives the transmission,
probability, $|t|^2$, depending in general on  the single particle
energy $\varepsilon$. Following ref.\cite{sh},  we can proceed to
the RG analysis which, in the limit of Strong Barrier, gives
 the conductance, $G$, as a function of  the temperature and $|t|$
 i.e.
\begin{equation} \label{gsp}
G\propto|t(\varepsilon, T)|^2\equiv |t(\varepsilon)|^2 T^{2
\alpha_{end}.}
\end{equation}
Here we introduced a second critical exponent,
\begin{equation} \label{alpend}
 \alpha_{end}=(1/K -1),
\end{equation}
also depending on $K$.

\

Experiments\cite{Postma01,Bozovic01} show transport through
an intrinsic quantum dot (QD) formed by a double barrier within a
1D electron system, allowing for the study of the  resonant or
sequential tunneling. The linear conductance  typically displays a
sequence of peaks, when the gate voltage, $V_g$, increases. Since
the initial theoretical work on this topic,\cite{kf,kf2,Furusaki}
the double-barrier problem in the absence of a magnetic field in a
TLl has attracted a significant amount of attention among
theorists.\cite{Sassetti95,Furusaki98,Braggio00,Thorwart02,Nazarov03,Polyakov03,Komnik03,Hugle04}
{The 1D nature of the correlated electrons is responsible for the
differences with respect to the quantum Coulomb blockade theory
for conventional, e.g., semiconducting QDs \cite{[13]}.  In the
(Uncorrelated) Sequential Tunneling (UST) approximation the
temperature dependence of the maxima of those peaks follows the
power law \cite{Furusaki98}
 \bea
 G_{max}\propto T^{\alpha_{end}-1},
\eea  with $\alpha_{end}$ being the DOS exponent for tunneling
into the end of a TLl. However, recent experiments \cite{Postma01}
suggest a different power law \bea \label{gct}
 G_{max}\propto T^{\alpha_{end-end}-1},
\eea with $\alpha_{end-end}=2\alpha_{end}$. This result follows
from the  Correlated Sequential Tunneling (CST) theory typical for
tunneling between the ends of two TLls.}

\

{\it Quasi 1D devices -}
Semiconductor QWs are quasi 1D devices  (having a width smaller than $1000
\AA$\cite{thor} and a length of some microns), where the electron
waves are in some ways analogous to electromagnetic waves in
waveguides. In these devices the electrons are confined to a
narrow 1D channel, with the motion perpendicular to the
channel quantum mechanically frozen out. Such wires can be
fabricated using modern semiconductor technologies, such as
electron beam lithography and cleaved edge overgrowth.

QWs are  usually made
  at the interface of different
thin semiconducting layers (typically $GaAs:AlGaAs$)
heterojunction, where a quasi two dimensional electron gas (2DEG)
can be formed by etching the
heterojunction 
\cite{thor}.

In  a recent experiment\cite{qw1} on long nanowires of degenerate
semiconductor $InSb$ (with a diameter  around $50 \AA$ and a length
of $0.1 - 1$ mm) a zero-field electrical conduction was observed,
over a temperature range $1.5 - 350 K$, as a power function of the
temperature with the typical exponent $\alpha_{Bulk} \approx 4$.
This value is about $10$ times larger than the one measured in
CNs, and the explanation of the  ratio $\alpha_{QW}/\alpha_{CN}\approx
10$  is the first result of this paper .

\

{\it Magnetic field effects - }
 Recently we already  discussed
the effects of a strong transverse magnetic field in both
QWs\cite{noiqw}, by focusing on the case of a very short range e-e
interaction, and large radius CNs\cite{noimf} for an unscreened
Coulomb interaction, by obtaining results in agreement with the
experimental data\cite{kanda}.  We explained this
behaviour\cite{noiqw,noimf} by discussing how
 the presence of a magnetic field produces the rescaling
of all repulsive terms of the interaction between electrons, with
a strong reduction of the backward scattering due to the edge
localization of the electrons.

\

{\it Impurities, QPCs and Intrinsic QD - }The magnetic induced
localization of the electrons should have some interesting effects
also on  the backward scattering, due to the presence of one or
more  obstacles along the QW, and hence on the corresponding
conductance, $G$\cite{noiqw}. Thus the main  focus of our paper is
to analyze   two barriers along a quasi 1D device
 (e.g.  two Quantum Point Contacts
(QPCs)\cite{thor}  at a fixed distance $d$ in a semiconductor
 QWs) forming
   an intrinsic QD,
 under the action of a transverse  magnetic field.
 QPCs  are constrictions defined in the plane of a 2DEG,
 with a width of the order of the electron Fermi
wavelength and a length much smaller than the elastic mean free
path. QPCs  proved to be very well suited for the study of
quantum transport phenomena. They have been realized in split-gate
devices, for example, which offer the possibility to tune the
effective width of the constriction, and thus the number of
occupied 1D levels, via the applied bias voltage.
 The presence of
a magnetic field in a QW interrupted by a QD  can have quite
interesting effects. In fact, in the ballistic regime,
 regular oscillations of $G(B)$ were measured\cite{vw} as a function of the increasing magnetic field,  and
 the presence of these peaks was discussed, as providing
evidence of an Aharonov-Bohm effect.


\

{\it Summary -} In this paper we want to discuss the  issues mentioned
above.

In section II we introduce a theoretical model which can describe
the QW under the effect of a transverse magnetic field, and we
discuss the properties of the interaction starting from the
unscreened long range Coulomb interaction in two dimensions.

In section III we evaluate the $bulk$ and $end$ critical exponents.
Then  we discuss the effects on them due to an increasing
transverse magnetic field. We remark that $\alpha_{bulk}$
characterizes the discussed power-law behavior of the TDOS, while
($\alpha_{end}$)  characterizes the temperature dependence of
$G_{max}$, in both the UST and the CST regime.
  Finally, we discuss the presence  of an
intrinsic QD formed by two QPCs, also by analyzing the
correspondence with the quantized magnetic flux linked together
with the current flowing in the cavity.

\section{Model and Interaction}

{\it Single particle - } A QW is usually defined by a parabolic
confining potential along one of the directions in the
plane\cite{me}: $V_W(x)=\frac{m_e}{2}\omega_d^2 x^2$. We also
consider a uniform magnetic field $B$ along the $\hat{z}$
direction and choose
the gauge  ${\bf A}=(0,Bx,0)$. 
In order to diagonalize the Hamiltonian for QWs, we introduce the
cyclotron frequency $\omega_c=\frac{eB}{m_e c}$ and the total
frequency $\omega_T=\sqrt{\omega_d^2+\omega_c^2}$, and we point out
that $p_y=v_y+eBx/(m_e c)$ commutes with the Hamiltonian
\begin{equation}\label{hnw2}
H =
\frac{\omega_d^2}{\omega_T^2}\frac{p_y^2}{2m_e}+\frac{p_x^2}{2m_e
}+\frac{m \omega_T^2}{2}(x-x_0)^2,
\end{equation}
where $x_0=\frac{\omega_c p_y}{\omega_T^2 m_e}$. The
diagonalization of  the Hamiltonian in eq.(\ref{hnw2}) yields two
terms: a quantized harmonic oscillator and a quadratic free
particle-like dispersion. This kind of factorization does not
reflect itself in the separation of the motion along each axis,
because the shift in the center of oscillations along $x$ depends
on the momentum $k_y$. Therefore, each electron in the system has a
definite single particle wave function \bea \label{wf}
\varphi_{n,k_y}(x,y)\propto e^{-\frac{\left(x-\gamma_\omega
k\right)^2}{2 \sigma_\omega^2}}h_{n}\left(x-\gamma_\omega
k\right)\frac{e^{i k_y y}}{\sqrt{2 \pi L_y}}, \eea where
$h_{n}\left(x\right)$ is the $n$-th Hermite polynomial,
$\gamma_\omega=\frac{\omega_c \hbar}{\omega_T^2 m_e}$ and
$\sigma_\omega=\sqrt{\frac{\hbar}{m_e \omega_T }}$. Now we are
ready to give a simple expression for the free electron energy,
depending on both the $y$ momentum $k$ and the chosen subband $n$
$$
\varepsilon_{n,k}=\frac{\omega_d^2}{2m_e \omega_T^2}\hbar^2
k^2+\hbar \omega_T(n+\frac{1}{2}),
$$
from which the magnetic dependence of the Fermi wavevector
follows
$$
k_F(\varepsilon_F,\omega_c)=\sqrt{\frac{2m_e \omega_T^2}{\hbar^2
\omega_d^2}\left(\varepsilon_F-\hbar
\omega_T(n+\frac{1}{2})\right)}.
$$

 Below we limit ourselves to electrons in a
single channel ($n=0$) and  calculate  a field-dependent
free Fermi velocity
\begin{equation}\label{vf}
v_F(\omega_c)=\frac{\omega_d^2}{m_e \omega_T^2}\hbar k_F\approx
\frac{\omega_d^2}{m_e \omega_c^2}\hbar k_F,
\end{equation}
where the approximation is valid for very strong fields.

\

{\it Electron-electron interaction - }
 {
In order to analyze in detail the role of the e-e interaction, we have
to point out that  quasi 1D devices have low-energy branches, at
the Fermi level, that introduce a number of different scattering
channels, depending on the location of the electron modes near the
Fermi points. It has been often discussed that processes which
change the chirality of the modes, as well as processes with large
momentum-transfer (known as backscattering and Umklapp processes),
are largely subdominant, with respect to those between currents of
like chirality (known as forward scattering
processes)\cite{8n,9n,egepj}.

}

 \

Now, following Egger and Gogolin\cite{egepj}, we introduce the
unscreened Coulomb interaction in two dimensions
 \bea \label{intx}
 V({\bf r}-{\bf r'})=\frac{c_0}{\sqrt{(x-x')^2+(y-y')^2}}. \eea Then, we can
calculate $U_0(k,\omega_c)$  starting from  the eigenfunctions
$u_{0,k_F}(x,y)$ and the potential in eq.(\ref{intx}).

\begin{figure}
\includegraphics*[width=1.0\linewidth]{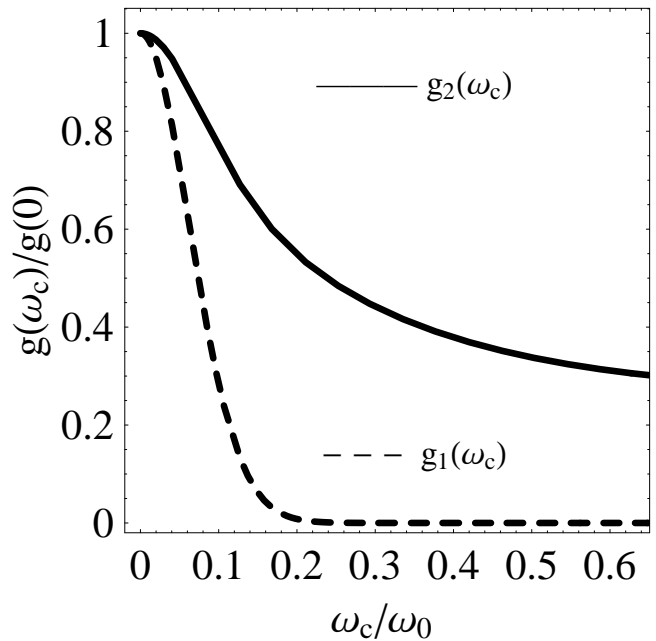}
\caption{{Scaling of the interaction with the magnetic field. The
forward scattering ($g_2$) term (solid line) is reduced by the
presence of the magnetic field ($\omega_c$), but this effect is
more consistent for the backscattering (dashed line) which
vanishes quickly with the increasing of $\omega_c$. Each value of
$g(\omega)$ is renormalized with respect to the corresponding value at zero
magnetic field ($\omega_c=0$). }}
\end{figure}
The fundamental interaction parameter is due to forward scattering
between opposite branches, corresponding to the interaction between
electrons with opposite momenta, $\pm k_F$, with a small momentum
transfer ($\sim q_c$). The strength of this term
$U_0(q_c,\omega_c)\equiv g_2$ is
\begin{eqnarray}
\label{uq}
U_0(q_c,\omega_c)& \approx & 2 U_0\\ \nonumber &\times&
\left( |\ln(\frac{q_c
\sigma_\omega}{4})|-\frac{\gamma_e}{2}-\frac{{{\gamma_\omega^2 k_F
}}^2}{{\sigma_\omega }^2} f(\frac{{{\gamma_\omega^2 k_F
}}^2}{{\sigma_\omega }^2})\right),
\end{eqnarray}
where $U_0$ is a constant parameter,
$\gamma_e$ is the Euler Gamma constant, $f$ is expressed
in terms of generalized hypergeometric functions 
.

 As we discussed above, the backscattering process, which
changes the chirality (with transferred momentum $2 k_F$), can be
neglected. This approximation becomes more suitable, when the
magnetic field increases, as  we show in Fig.(1).

\section{Results}

{\it The bulk and the end critical exponents -}   The
$\alpha_{Bulk}$ in a QW has to be $10$ times larger than in a CN
(i.e.  $\alpha_{QW}\approx 4$) and this  is due to a difference in
the Fermi velocity. Let us recall that a typical Single Wall CN,
with a longitudinal length $L_{CN} \approx 3-10 \mu m$ and a
radius $R_{CN}=1.38 nm$, has critical exponent $\alpha\approx
0.3-0.4$ corresponding to $g_2^{CN}\approx 1-1.5 \times 10^2
v_{CN}$, where $v_{CN}=8\times 10^5 m/s$ is the Fermi velocity in
a CN, as it can be obtained by applying eq.(\ref{al1}) ($K \approx
0.18$ for a  Single Wall CN of length $3\mu m$\cite{9n}). For a
comparison of our model for a QW with the related measurements, we
have to calculate the frequency $\omega_d$ starting  from the
width, $R$, of the QW as $\omega_d \approx \frac{\hbar
(2\pi)^2}{m_0 R^2}$, where we have to consider the effective mass
($m_0=0.067 m_e$ for $AsGaAs$).

A semiconducting QW made in $AsGaAs$ 2DEG has typically a length
$L\sim 10-100 \mu m$ and a width $20-30 nm$. Thus we obtain $\hbar
\omega_d \approx 20-40 meV$. The Fermi velocity can be obtained,
after introducing the Fermi wavevector $k_F$
 corresponding to a half filled subband, as 
$v_F\approx  10^{3}- 10^{4}m/s$.

Now we consider the QW in ref.\cite{qw1} with  a longitudinal
length $L=0.1 mm$ and transverse size $R=5 nm$. Because the
strength of the e-e interaction, $g_2$, depends on the logarithm
of the ratio between the transverse and the longitudinal
dimensions, we can conclude that it  is rather the same for this
QW and a typical Multi Wall CN ($g_2^{QW}\approx g_2^{CN}$). However, the
large difference between the corresponding Fermi velocities (a
factor $\sim 10^3$) yields strong effects on the ratio $g_2/v_F$.
%
From the introduction of the experimental parameters for the QW in
ref.\cite{qw1} it follows $\alpha_{bulk}\approx 3-4$, in good
agreement with experimental results and more then $10$ times
larger than the one measured in CNs.

However,  by introducing the expression from eq.(\ref{uq}) into
eq.(\ref{al1}), it follows that the bulk critical exponent is
reduced by the presence of a magnetic field, as we show in
Fig.(2).
\begin{figure}
\includegraphics*[width=1.0\linewidth]{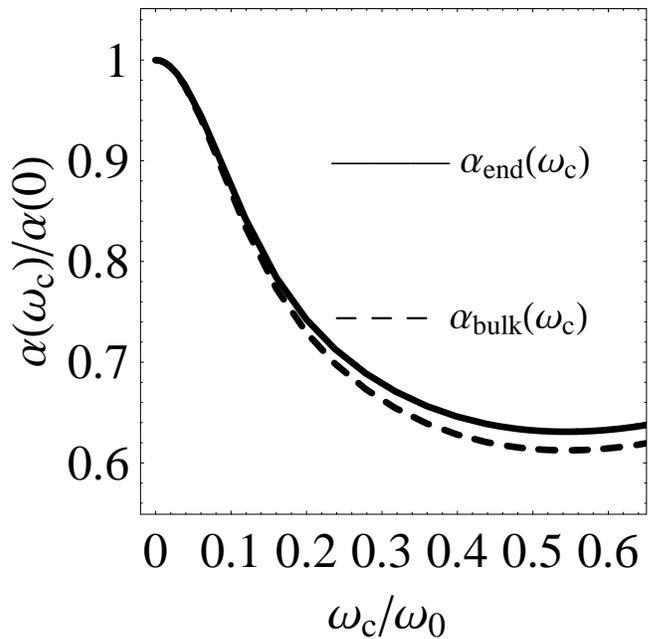}
\caption{{Critical exponents versus magnetic field for a QW:
$\alpha_{Bulk}$ is calculated following eq.(\ref{al1});
$\alpha_{end}$ is calculated following eq.(\ref{alpend}). The
magnetic field yields  a strong reduction of both critical
exponents. We can consider $\omega_c/\omega_d=0.5$ corresponding
to $B\approx 0.5$ T for the QW in the experiment of ref.\cite{qw1}
}}
\end{figure}
 We can conclude that  the magnetic field alters the bulk
exponent: on the one hand, the localization of the edge states is
responsible for the reduction of $\alpha_{bulk}$, because of the
attenuation of the forward scattering between opposite branches;
on the other hand,  also the Fermi velocity is renormalized, as
shown in eq.(\ref{vf}). This  prediction can be extended to
$\alpha_{end}$, calculated following eq.(\ref{alpend}), as we show
in Fig.(2).

\

{\it The intrinsic Quantum Dot  -} {When there are some obstacles
to the free path of the electrons along a 1D device (e.g.  QPCs,
which shrink the width of a QW), a scattering potential has to be
introduced in the theoretical model. Details about calculations
concerning the presence of obstacles in a 1D electron systems
were discussed in refs.(\cite{kf2,fn}), where the problem is
mapped onto an effective field theory using bosonization and then
approached using a RG analysis. The presence of two barriers
along a QW at a distance $d$  can be represented by a potential
$$V_B(y)=U_B\left(f(y+\frac{d}{2})+f(y-\frac{d}{2})\right),$$ where
$f(y)$ is a square barrier function, a Dirac Delta function or any
other function localized near $y=0$. In general we can  analyze
the single particle transmission in the presence of a magnetic
field, $t(\varepsilon_F,B)$,  by identifying the off-resonance
condition ($|t|=0$), where electrons are strongly backscattered by
the barriers, and the on-resonance condition ($|t|=1$), where  the
scattering at low temperatures is negligible.

 \

Now we can discuss  some details of the results obtained for a double
square barrier: the magnetic dependence of the peaks in the
transmission is shown in Fig.(3.top), where we report the
transmission $T=|t|^2$ versus $\omega_c$, which exhibits
 a magnetically tuned transport through the QW. In particular, assuming
that there are two identical, weakly scattering barrier at a
distance $d$, the transmission is non-zero for particular values of
$k_F$, so that $\cos(k_F d)\approx 0$.
\begin{figure}
\includegraphics*[width=1.0\linewidth]{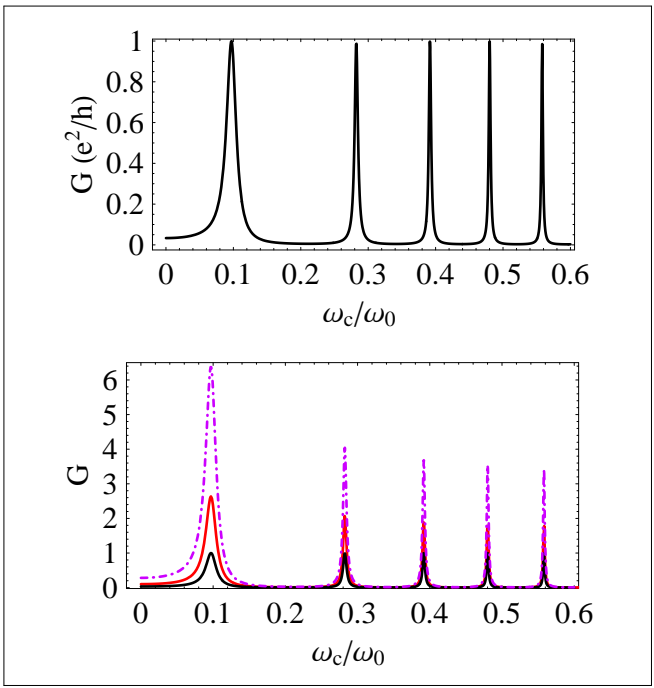}
\caption{{(Top) Ballistic conductance as a function of the
magnetic field: the use of a double square barrier model, for the
cavity formed by two QPCs, allows for the exploration of the
backward scattering oscillations due to the magnetic field. We
observe the appearing of resonance peaks, as a function of the
magnetic field. (Bottom) Magnetic field  dependent conductance
obtained for $3$ different  values of the temperature, following
the CST theory (the solid line corresponds to the top panel  with
the lowest temperature $T_0$ while the dashed line corresponds to
a higher value of $T\approx 1.2 T_0 $). The observed reduction of
the peaks height corresponds to the predicted reduction of
$\alpha_{end}$ with $B$. }}
\end{figure}

\

It is quite interesting to analyze the correspondence between the
resonance peaks in the conductance and the geometry of the current
vector field (see Fig.(4)) for strong magnetic fields, starting
from the usual resonance condition: the $n-th$ peak corresponds to
$k_F d \approx n \pi$.

Some papers about the ballistic transport in the presence of a
magnetic field\cite{vw} discuss the presence of the
conductance peaks, interpreting it as the evidence of an Aharonov-Bohm effect. We
can explain that, by considering the localization of the edge
states in the wire ($\langle x\rangle=\pm \gamma_\omega k_F$), so
that the path of the electrons encloses a surface $S= 2
\gamma_\omega k_F d$. In the limit of strong $B$ ($\gamma_\omega
\approx (\hbar)/(m_e \omega_c)$), we obtain a value for the flux of
$B$ in the presence of a transmission peak
$$
\Phi_S(B)\approx B  (2 \gamma_\omega k_F d) \approx \frac{2 c
\hbar}{e} k_F d \approx n \frac{2 c h}{e}=n \Phi_0.
$$
\begin{figure}
\includegraphics*[width=1.0\linewidth]{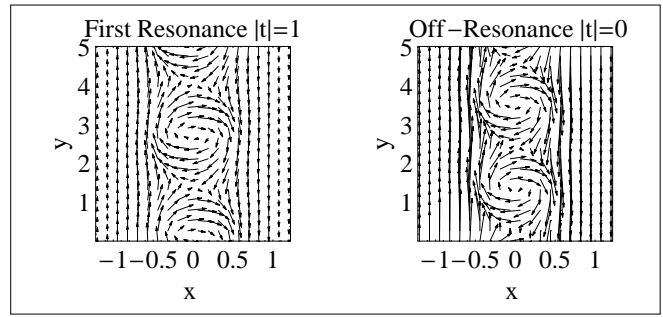}
\caption{{ The current vector field between the barriers, in the
resonance and the off resonance conditions, for two square barriers, in
the presence of a magnetic field. }}
\end{figure}
This can also be seen by analyzing the presence of an integer
number of "circles of current", between the two barriers, in
correspondence to the peaks (see Fig.(4.left)). In Fig.(4.right)
we show the off resonance behaviour corresponding to $|t|=0$. In
both Figs. (4) each circle of current represents one electron
which brings a quantum of magnetic flux $\Phi_0$.

Following the theoretical approach to the TLl in the presence of
two barriers\cite{kf,kf2,Furusaki,Furusaki98}, we can calculate
the {\em resonant scattering} condition, which can   give rise to
perfect transmission even for $K<1$. It   corresponds to an
average particle number between the two barriers of the form
$\nu+1/2$, with integer $\nu$, i.e. the QD is in a degenerate
state. If interactions between the electrons in the QD are
included, one can recover the physics of the Coulomb
blockade\cite{kf,kf2,Furusaki,Furusaki98}. The main difference is
due to the temperature dependence of the conductance $G_{max}$. For
the calculation of this dependence we can follow the CST mechanism
recently proposed\cite{Postma01}, in order to explain
 the unconventional power-law
dependencies in the measured transport properties of a CN. In this
theory the  electrons tunnel coherently, from the end of one CN
lead to the end of the other CN lead, through a quantum state in
the island. In this picture, the island should be regarded as a
single impurity.   The power law dependence of the conductance due
to this tunneling mechanism is reported in eq.(\ref{gct}) and is
shown in Fig.(3.bottom).

\section{Conclusions }
In this paper we showed how the presence of a magnetic field
modifies the role played by both the e-e interaction and the
presence of obstacles in a QW.

The first prediction that comes from our study is that  the bulk
critical exponent $\alpha_{bulk}$ of a semiconductor QW should  be
10 times larger than the one measured in a typical CN according to
the experimental results.
 We also predict a significant reduction  of both critical exponents
as the
magnetic field is increased.
The magnetic dependent value of $\alpha_{end}$ determines the
temperature dependent $G$ in the sequential tunneling regime.

Our second prediction concerns the presence of some peaks in the
small bias conductance versus magnetic field when  two QPCs in are
put  series along  a QW,  forming an intrinsic QD (see Fig.(3)).
The presence of magnetic field dependent peaks in the transmission
can be used, in order to construct a "magnetic field transistor"
also in the temperature regime corresponding to the Luttinger
liquid (a room temperature transistor, if we look at the device
proposed in ref(\cite{Postma01})). Thus we take into account a
semiconducting QW made in $AsGaAs$ 2DEG, which  typically have a
length $L\sim 10-100 \mu m$ and a width $20-30 nm$, i.e. $\hbar
\omega_d \approx 50-100 meV$. The corresponding magnetic energy
$\hbar \omega_c/B\approx 15 meV/T$ is comparable with the
confining one $\hbar \omega_d$, while the strong renormalization
of the effective electron mass reduces by a factor $100$ the
Zeeman spin splitting. If we fix two QPCs at a distance $d\approx
200-250 nm$,
we predict that some peaks (about $5-10$) have to be observed in
the conductance, for values of the magnetic field between $0$ and
$4$ T. The effects of very strong magnetic fields (much larger
than those considered here) can dramatically change the behaviour
of the system, as we showed in our previous paper\cite{noiqw},
where we discussed also the spin polarization in QWs.

Our third result concerns  the two different explanation for  the
peaks in the conductance. The discussed on-resonance condition in
the TLl approach  corresponds to the presence of an average
particle number, $\nu$, in the cavity formed by the QPCs. On the
contrary, the presence of a quantized circulating current,
corresponding to the conductance peaks, was read as providing
evidence of an Aharonov-Bohm effect in the ballistic regime. Thus
we suggest that each electron in the QD has to bring a magnetic
flux quantum.


\bibliographystyle{prsty} 
\bibliography{}

\end{document}